\documentclass [aps,showpacs,showkeys,preprint,amsmath,amssymb]{revtex4}

\usepackage{graphicx}
\usepackage{amsmath}

\begin{document}

\title{ \bf Interfacial effects on the polarization of  $BiFeO_{3}$ films}
\author{\bf  Yin-Zhong Wu$^{a,b}$, Tao Pan$^b$, and Zhen-Ya Li$^c$}
\affiliation{$^{a}$Jiangsu Laboratory of Advanced Functional materials, Changshu Institute of Technology, Changshu 215500, China
\footnote{Email: yzwu@cslg.edu.cn}\\
$^{b}$school of science, Suzhou University of Science and Technology, Suzhou 215011, China\\
 $^{c}$Physics Department and Jiangsu key laboratory of thin films, Soochow University, Suzhou 215006, China }

\begin{abstract}
 By considering an interfacial layer between the electrode and the $BiFeO_{3}$($BFO$) layer,
 the polarization and the hysteresis behavior of $BFO$ film are simulated. It is found that the non-ferroelectric interface will increase the
 coercive field, and remarkably suppress the polarization of the ultrathin film under low applied fields.
 Due to the competition between the interfacial effect and the internal compressive stress, the maximum polarization on the P-E loop of a $BFO$ film
 can be independent on the film thickness under an adequate applied field.
\end{abstract}

\pacs{77.80.Dj} \keywords{A. ferroelectrics; A. thin film; A. surfaces and interfaces}

\maketitle
\section{Introduction}

Recently, the perovskite-type oxides, which display ferroelectric and magnetic properties, have attracted considerable interest for their
potential applications in multifunctional magnetoelectric devices\cite{0}. It has been shown that $BFO$ is a good candidate which exhibits the
coexistence of ferroelectric and antiferromagnetic orders with high Curie temperature\cite{t1} and N$\acute{e}$el temperature\cite{t2}. Numerous
studies have been performed on $BFO$ samples and especially, more recently, on thin films. Generally, bottom and top electrodes are fabricated
to form a capacitor configuration for ferroelectric measurements. The physical origin of the interfacial layer has been proposed to be
process-induced domain evolution or the effect of lattice misfit between the electrode and surface of the film. The interfaces in $BFO$ film
capacitors have already been verified by using X-ray diffraction and scanning transmission electron microscope high-angle annular dark-field
imaging\cite{5}, and the effect of the interface on the dielectric behavior of $BFO$ film capacitor was also investigated in experiment\cite{4}.
For epitaxial system, the substrate-induced strains have been proven to have great influence on the ferroelectric properties of ferroelectric
 material\cite{fe-strain}.
The previous studies show that substrate-induced strains can lead to a substantial increase in spontaneous polarization for BFO
films\cite{7,8,9}. The mechanical substrate effects and thickness dependence of ferroelectric properties in epitaxial BFO films are investigated
within the framework of Landau-Devonshire theory.
 Although the interface between the electrode and the BFO layer has been reported in experiment\cite{5,new}, there is no theoretical investigation of the
 interfacial effect on the polarization of BFO film. In this paper, the interfaces between electrode and $BFO$ layer are considered with different film thickness. It is found that the interface
will have great influence on the polarization and the hysteresis behavior of ultrathin $BFO$ films. Due to the interfacial effect, the maximum
polarization on the P-E loop of thin film can be lower(or higher) than that of thickness film, and the maximum polarization can also be
thickness-independent through the competition between the interfacial effect and the compressive stress under an adequate applied field.

\section{Model and Theory}

 Here, we consider a $BFO$ film grown on $SrTiO_{3}$ substrate with a bottom
$SrRuO_{3}$ electrode and a top $Pt$ electrode\cite{33,6}. It is known that there is a clear and sharp interface between $BFO$ layer and the
bottom $SrRuO_{3}$ electrode, and a rough, inter-diffused interface between $BFO$ layer and the top $Pt$ electrode, which can be confirmed by
the cross-section HR-TEM image\cite{5,new}. It is reasonable that the interface between BFO and the bottom electrode can be neglected for
simplicity. Therefore, we simulate the $BFO$ film by a ferroelectric layer with a top non-switching interface, and investigate the effects of
the interface on the P-E loop with different film thickness. The electric displacement across BFO layer is given by
$D_{f}=\epsilon_{f}E_{f}+P_{f}$, where $\epsilon_{f}$, $E_{f}$ and $P_{f}$ denote the permittivity, the local electric field and the spontaneous
polarization of BFO layer, respectively. The electric displacement across the interface is $D_{in}=\epsilon_{in}E_{in}$, where $\epsilon_{in}$
and $E_{in}$ stand for the permittivity and the local electric field of interface. The applied field $E_{ex}$ can be written as
\begin{equation}
E_{ex}=(1-v_{in})E_{f}+v_{in}E_{in},
\end{equation}
where $v_{in}$ is the thickness ratio of interface. The continuity of the total current density across the film can be expressed as
\begin{equation}
J=\sigma_{in}E_{in}+\epsilon_{in}\dot{E}_{in}=\sigma_{f}E_{f}+(\epsilon_{f}+\frac{\partial P_{f}}{\partial E_{f}}) \dot{E}_{f}.
\end{equation}
Combining Eq.~(1) with Eq.~(2), we obtain
\begin{eqnarray}
&&\sigma_{in}E_{ex}+\epsilon_{in}\dot{E}_{ex}=[(1-v)\sigma_{in}+v\sigma_{f}]E_{f}\nonumber\\
&&+[v(\epsilon_{f}+\chi_{f})+\epsilon_{in}(1-v)]\dot{E}_{f},
\end {eqnarray}
where $\sigma_{in}$ and $\sigma_{f}$ are the conductivities of top interface and $BFO$ layer, respectively, $\chi_{f}$ is the electric
susceptibility of $BFO$ layer, which equals $\partial{P_{f}}/\partial{E_{f}}$\cite{1}
\begin{eqnarray}
\chi_{f}&=&P_{f}\{2\delta cosh^2[(E_{f}-E_{c})/2\delta]\}^{-1},\\
\delta&=&E_{c}[log(\frac{1+P_{fr}/P_{fs}}{1-P_{fr}/P_{fs}})]^{-1},\nonumber
\end {eqnarray}
where $P_{fr}$, $P_{fs}$ and $E_{c}$ denote the remanent polarization, saturate polarization and coercive field of $BFO$ layer, respectively,
which can be determined through Landau-Devonshire theory\cite{8}.

The $BFO$ films with different thickness are supposed to grow on (001)-oriented $SrTiO_{3}$ substrate. A tetragonal structure of $BFO$ thin film
based on cubic perovskite structure with symmetry lowered to P4mm(neglect the small monoclinic distortion of about $0.5^{o}$ of c axis) is
suggested with an elongated c axis\cite{w2}. The polarization for such a single phase is along with $c$ axis. The thermodynamic potential of
pseudo cubic $BFO$ film is expressed as the function of magnetization $M_{f}$, polarization $P_{f}$, temperature $T$ and misfit strain
$u_{m}=(a_{s}-a_{0})/a_{s}$, where $a_{s}$ is the in-plane lattice parameter for substrate and $a_{0}$ is the equivalent cubic lattice constant
of free standing film. We do not include the magnetic effect and the coupling between magnetic order and ferroelectric order, owing to a
relative weak magnetic order compared to the stronger ferroelectric one. The thermodynamic potential of $BFO$ layer can be written as following,
\begin{equation}
\tilde{F}(P_{f},T)=\alpha_{1}^{*}P^{2}_{f}+\alpha_{11}^{*}P^{4}_{f}+\alpha_{111}P^{6}_{f}-E_{f}P_{f}+\frac{u_{m}^{2}}{S_{11}+S_{12}},
\end{equation}
where $\alpha_{1}^{*}=\alpha_{1}+u_{m}\frac{2Q_{12}^{E}}{S_{11}+S_{12}}$, $\alpha_{11}^{*}=\alpha_{11}+\frac{(Q_{12}^{E})^{2}}{S_{11}+S_{12}}$,
$Q_{ij}^{E}$ is the electrostricitive coefficient, $S_{ij}$ is the elastic compliances, and the dielectric stiffness coefficient
$\alpha_{1}=(T-Tc)/2\epsilon_{0}C$. The equilibrium state polarization $P_{f}$ under a local field $E_{f}$, $P_{fr}$ and $P_{fs}$ in Eq.~(4) can
be numerically resolved from Eq.~(5) by $\partial \tilde{F}/\partial P_{f}=0$,
\begin{equation}
2\alpha_{1}^{*}P_{f}+4\alpha_{11}^{*}P_{f}^{3}+6\alpha_{111}p_{f}^{5}-E_{f}=0.
\end {equation}
In order to introduce the misfit strain into the epitaxial $BFO$ film, an effective substrate lattice parameter $\tilde{a}_{s}$ is defined as
$\tilde{a}_{s}=\frac{a_{s}(T)}{\rho a_{s}(T)+1}$, which is used to calculate the misfit strain $u_{m}$ instead of actual substrate lattice
parameter $a_{s}$, and $u_{m}=\frac{\tilde{a}_{s}(T)-a_{0}(T)}{\tilde{a}_{s}(T)}$. $\rho$ is the equilibrium linear misfit dislocation density
at the deposition temperature $T_{G}$. $\rho=\frac{u_{m}(T_{G})}{a_{0}(T_{G}}(1-\frac{h_{p}}{h})$, where $h_{p}$ is the critical film thickness
corresponding to the generation of misfit dislocation. All the coefficients in Landau expansion used in this letter are given in Ref.~\cite{11}.

\section{Numerical Results and Discussions}

 From Eq.~(6), we calculate the spontaneous polarization $P_{f}$ without considering the interface.
The hysteresis behavior of the single $BFO$ layer is shown in Fig.~1, where solid line stands for the hysteresis loop of the 77-nm-thick film,
and dotted line for the 960-nm-thick film.
\begin{figure}
\includegraphics[width=2.5in]{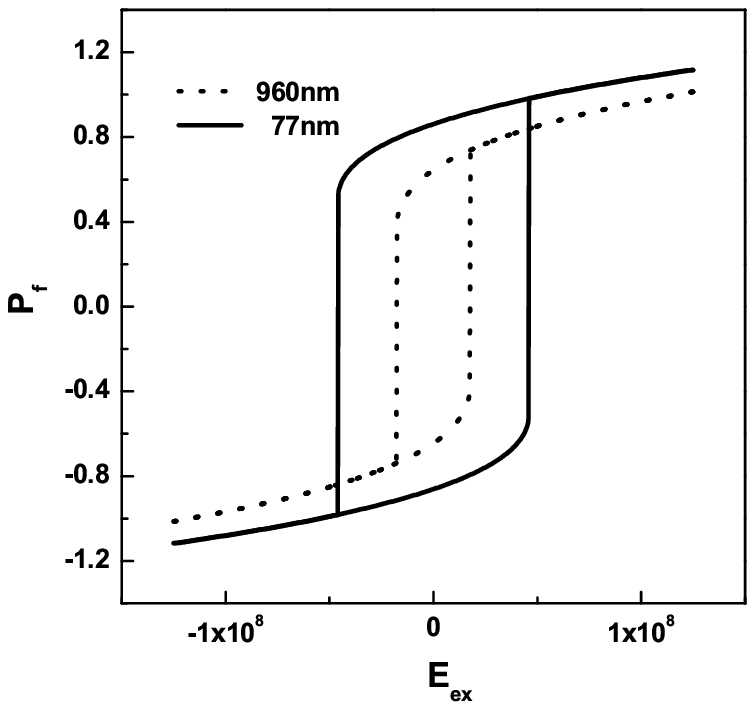}
\end{figure}
The remanent polarization of the $960$nm $BFO$ film is $64\mu C/cm^{2}$, which is close to the polarization of single crystal\cite{last}. We can
see that the spontaneous polarization increases as the thickness decreases, and our results are qualitatively consistent with previous
theoretical works\cite{7,8}. It is obviously that the compressive stress between $BFO$ layer and substrate $SrTiO_{3}$ will remarkable increase
the polarization, and the spontaneous polarization $P_{f}$ will increase with the decrease of thickness. However, as observed by AFM\cite{5} and
simulated by the capacitor configuration\cite{4}, there really exists an interface between top electrode and $BFO$ layer. So, the measured
polarization is not the polarization of a single ferroelectric layer. In the following, an interface between the top $Pt$ electrode and $BFO$
layer is introduced, and two $BFO$ film samples of thickness $77$nm and $960$nm are selected in our simulation\cite{7}. The dielectric constant
and the conductivity of the $BFO$ layer are selected as 100\cite{w5} and $1\times 10^{-10}\Omega^{-1}cm^{-1}$\cite{33}, respectively. Due to the
interface is formed between the metal material and $BFO$ layer, we assume its conductivity and dielectric constant take the values of $1\times
10^{-9}\Omega^{-1}cm^{-1}$\cite{4} and $20$. The thickness of interface is introduced by taking into account the thickness ratio of interface to
$BFO$ layer. The ratio $v$ will increase as the decrease of film thickness, and the ratios $v$ are taken as $0.02$ and $0.002$ for the two
samples.

  For a given sinusoidal field $E_{ex}$, the effective field $E_{f}$ can be calculated from Eq.~(3) as a
  function of time $t$. Then the polarization $P_{f}$ at the $m$th increment in time is calculated by using $P_{f_{m}}=P_{f_{m-1}}+(E_{f_{m}}-E_{f_{m-1}})
  \frac{\partial P_{f}}{\partial E_{f}}|_{m-1}$. The simulated hysteresis loops corresponding to different applied fields
  are shown in Fig.~2, the frequency of the applied field in our simulation is $2$kHz. The solid lines in Fig.~2 denote the loops of
  77-nm-thick film, and the dotted lines stand for the loops of 960-nm-thick film. From Fig.~2(a) to (d), the maximum field
  are selected as $1.0$ MV/cm, $1.25$ MV/cm, $1.6$ MV/cm and $2.5$ MV/cm, respectively. By taking into account the interface, the local field
  $E_{f}$ will decrease as the decrease of thickness under the same applied field. For the 960-nm-thick film, due to
the small proportion of the interface component, the local field $E_{f}$ is close to the applied field, and the P-E loop is already saturated in
Fig.~2(a), whereas the P-E loop for 77-nm-thick film does not saturated, and the maximum polarization on the P-E loop is smaller than the
saturated polarization of 960-nm-thick film. One can see, from Fig.~1 and Fig.~2(a), that the interface will great suppress the polarization of
the ultrathin film, and the interfacial effect of an ultrathin film is more important than the internal stress effect under a low applied field.
For a selected applied field $E_{max}$=$1.25$ MV/cm in Fig.~2(b), the maximum polarization on the P-E loop is independent on the thickness of
the film, which is the consequence of the competition between the interfacial effect and the internal compressive stress effect. In Fig.~2(c),
the maximum polarization of the ultrathin film is larger than that of the thick film, which had also been observed in experiment(See Fig.~4(d)
in Ref.~\cite{w3}), and it shows that the interfacial effect is less dominant for the ultrathin film under a high applied field. As further
increasing the maximum field, the polarizations for the two samples are both saturated, and the saturated polarization of the 77-nm-thick film
is larger than that of the 960-nm-thick film. In addition, the P-E loops for materials such as $BFO$ film are often dominated by leakage
current, and the saturated P-E curve in Fig.~2(d) for an ultrathin film can only be observed for samples with high resistivity and high
breakdown field.
\begin{figure}
\includegraphics[width=4.0in]{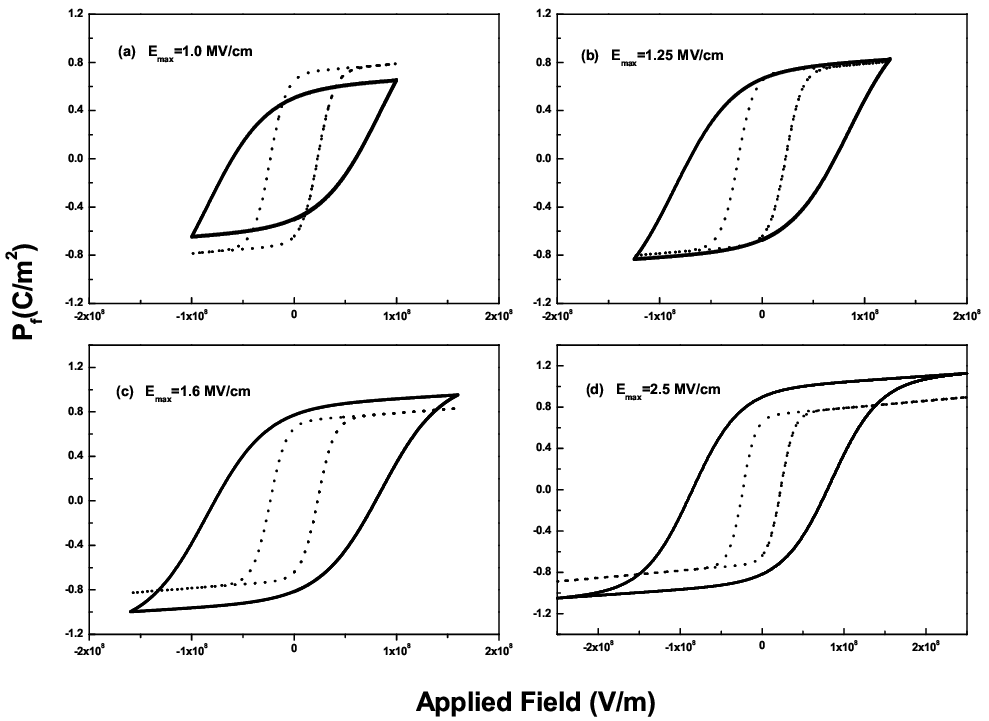}
\end{figure}

In summary, the epitaxial $Pt/BFO/SrRuO_{3}/SrTiO_{3}$ film is simulated by a multilayer model which takes the effect of interface. It is found
that the existence of interfacial layer will increase the coercive field, and change the shape of the P-E loop, especially for an ultrathin
film. The P-E loop will become more slant for the ultrathin film, which is resulted from the decreasing of the effective field in $BFO$ layer.
For different maximum fields, the maximum polarization of the ultrathin film can be smaller(or larger) than that of the thick film. For an
adequate applied field, the maximum polarization on the P-E loop could be thickness-independent, and the origin is caused by the competition
between the interfacial effect and the internal compressive stress. It is concluded that the effect of interface between the electrode and $BFO$
layer has great influence on the polarization of $BFO$ film. When measuring the saturated P-E loops for multiferroic ultrathin films with metal
contacts, one should apply a field large enough to saturate the polarization.
\begin{acknowledgments}
This work was supported by National Science Foundation of China(Grant No.10874021 and 10774107), and Science Foundation of Education Committee
of Jiangsu Province(Grant No.~07KJB140002).
\end{acknowledgments}

\vspace{2cm}
Figure Captions\\

FIG1:  Polarization of $BFO$ layer as a function of the applied field without considering the interface.\\

FIG2: P-E loop of $BFO$ film with interface for different applied fields. Solid lines correspond to the 77-nm-thick film, and the dotted lines
correspond to the 960-nm-thick film.


\begin{thebibliography}{99}
\bibitem{11}List of parameters in the calculation of spontaneous polarization in Eq.~(6) are selected as\cite{8}.
$T_{C}=830^{o}C$, T=77K, $C=2.45525\times {10^{7}}^{o}C$, $\alpha_{11}=6.4\times 10^{6}m^{5}/C^{2}F$, $\alpha_{111}=6\times 10^{5}m^{9}/C^{4}F$,
$S_{11}=4.35\times 10^{-12}m^{2}/N$, $S_{12}=-4\times10^{-12}m^{2}/N$, $a_{s}(T)=3.905$, $a_{0}(T)=3.962$, $TEC=10.0\times10^{-6}K^{-1}$ and
$h_{p}=10nm$ for BFO thin film, and $TEC=11.7\times10^{-6}K^{-1}$ for $SrTiO_{3}$ substrate.
\end{thebibliography}
\end{document}